\documentclass{article}
\usepackage{arxiv}
\usepackage[utf8]{inputenc} 
\usepackage[T1]{fontenc}    
\usepackage{hyperref}       
\usepackage{url}            
\usepackage{booktabs}       
\usepackage{amsfonts}       
\usepackage{nicefrac}       
\usepackage{microtype}      
\usepackage{lipsum}		
\usepackage{graphicx}
\usepackage{doi}
\usepackage[numbers]{natbib}

\title{On the validity of fMRI studies with subject-level data processed through different pipelines}
\author{
  Elodie Germani \thanks{Co-first authors}\\
  Univ Rennes, Inria, CNRS, Inserm \\
  Rennes, France\\
  \texttt{elodie.germani@irisa.fr} \\
  \href{https://orcid.org/0000-0002-5786-9538}{\includegraphics[scale=0.06]{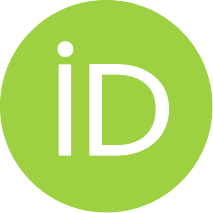}\hspace{1mm} ORCID: 0000-0002-5786-9538} \\
   \And
  Xavier Rolland \footnote[1]{}\\
  Univ Rennes, Inria, CNRS, Inserm \\
  Rennes, France\\
  \texttt{xavier.rolland@ac-grenoble.fr} \\
    \And
  Pierre Maurel \footnote[2]{} \\
  Univ Rennes, Inria, CNRS, Inserm \\
  Rennes, France\\
  \href{https://orcid.org/0000-0001-5988-5446}{\includegraphics[scale=0.06]{orcid.pdf}\hspace{1mm} ORCID: 0000-0001-5988-5446} \\
   \And
  Camille Maumet \thanks{Joint senior authorship}\\
  Univ Rennes, Inria, CNRS, Inserm \\
  Rennes, France\\
  \href{https://orcid.org/0000-0002-6290-553X}{\includegraphics[scale=0.06]{orcid.pdf}\hspace{1mm} ORCID: 0000-0002-6290-553X}
}

\date{}

\renewcommand{\shorttitle}{Validity of between-group fMRI studies}

\hypersetup{
pdftitle={fMRI compatibility},
pdfsubject={q-bio.NC, q-bio.QM},
pdfauthor={Elodie Germani, Xavier Rolland, Pierre Maurel, Camille Maumet},
pdfkeywords={First keyword, Second keyword, More},
}

\begin{document}
\maketitle
\bibliographystyle{unsrtnat}

\begin{abstract}
In recent years, the lack of reproducibility of research findings has become an important source of concerns in many scientific fields, including functional Magnetic Resonance Imaging (fMRI). The low statistical power often observed in fMRI studies was identified as one of the leading causes of irreproducibility. The development of data sharing opens up new opportunities to achieve larger sample sizes by reusing existing data. FMRI studies proceed by first preparing subject-level data using a given analysis pipeline and then combining those into a group analysis. Historically the subject-level analysis pipeline was identical for all subjects. As practices evolve towards more data reuse, researchers might want to directly combine subject-level data thata were processed using different pipelines. Here, we investigate the impact of combining subject-level data processed with different pipelines in between-group fMRI studies. 

We used the HCP Young-Adult dataset (N=1,080 participants) processed with 24 different pipelines. We then performed between-group analyses comparing subject data processed with different pipelines. We worked under the null hypothesis of no differences between subjects and compared the estimated false-positive rates obtained with the nominal rates. We showed that the analytical variability induced by the parameters explored in this dataset increases the false positive rates of studies combining data from different pipelines. We conclude that different processed subject data cannot be combined without taking into account the processing applied on these data.
\end{abstract}

\keywords{neuroimaging, analytical variability, pipelines, validity, data re-use}

\section{Introduction}
Task-based functional Magnetic Resonance Imaging (fMRI) is a neuroimaging technique that studies the activation of brain regions during specific experimental protocols. The main form of fMRI is Blood-Oxygen Level Dependent (BOLD) fMRI, which uses MRI to measure, at every position of the brain, a BOLD signal whose variations in time are related to brain activity. In fMRI studies, the raw data resulting from an MRI session for a given subject consists in a temporal sequence of 3D brain images. After acquisition, these data are preprocessed and analyzed at the subject and group-level to obtain 3-dimensional statistic maps representing the activation of the brain for the paradigm of interest.

Although fMRI has been a useful tool to provide information about the roles of the different regions of the brain, multiple concerns have been raised over the last few years regarding research practices that might impact the reproducibility of neuroimaging studies~\cite{bowring2019exploring,botvinik2020variability,li2021moving}. One of those factors is the overall low statistical power of fMRI studies: having low sample sizes makes it harder to find true positive results, hence increasing the likelihood that any positive result is false~\cite{button2013power}. In 2015, the median sample size in fMRI studies was at almost 30 subjects for single-group studies~\cite{poldrack2017scanning}. While the sample sizes in fMRI studies have been increasing over the past few years with the development of large-scale studies~\cite{sudlow_uk_2015,van2013wu}, there is still a crucial effort that has to be done in order to increase statistical power.
There are multiple ways to overcome this lack of reproducibility induced by low sample sizes. For example, meta-analyses can be used to combine results from multiple studies (group-level statistical maps in our case) to observe converging results~\cite{salimi2009meta}. However, there are several limitations to this method, notably due to the existence of publication bias~\cite{ioannidis2014publication}. Another possibility is to take advantage of data sharing: today, with the importance given to open science~\cite{poldrack2014making} and the development of research infrastructures~\cite{poldrack2013toward,gorgolewski2015neurovault}, more and more neuroimaging data from various studies are made available to the scientific community. These datasets include subject data from multiple different studies, which can be used and combined in new studies. Being able to combine such data would allow us to have larger sample sizes per studies, and thus more robust results.
Although most studies based on data reuse currently focus on raw data, we expect that in the future more and more processed datasets will be made available. Sharing of processed data may appear because of privacy reasons, and also in order to avoid having to perform the subject-level processing each time someone wants to reuse the subject data. FMRI studies require multiple processing steps on the data, both at the subject-level (preprocessing of the raw fMRI data to prepare these for statistical analysis, and first-level analysis for each subject) and at the group-level (second-level statistical analysis using the subject data resulting from first-level analysis).
Within a study, there are many factors of variability which may have an effect on the analysis. Some of these are unwanted factors of variability on which the results of a study may depend. Sources of variability include acquisition instruments, acquisition protocol, and differences in the processing and analysis protocol~\cite{friedman2006reducing}. Here, we focus on the variability resulting from the processing and analysis protocol used on the data, which is also known as analytical variability. This type of variability has been studied in various cases, for instance the variability related to software and software versions in~\cite{bowring2019exploring, gronenschild2012effects}. At each processing step within fMRI studies, multiple methodological choices are available: performing certain operations or not, the order of these operations, parameter values for a specific operation, or using a specific software package that may implement a different algorithm from the others to perform an operation~\cite{carp2012plurality}. The series of operations that will be performed on the data is called a ``pipeline". FMRI pipelines represents the whole analysis, but they can be split into different subparts: for example, subject-level pipelines only perform preprocessing and first-level analysis on the subject data.

The variety of possible methodological choices leads to a large amount of possible pipelines for a same analysis. Multiple studies~\cite{ganz2020false,gronenschild2012effects,botvinik2020variability,bowring2019exploring,carp2012plurality,glatard2015reproducibility} have shown that variability in analytical choices in neuroimaging studies can lead to differences in results when doing the same experience with different analysis methods. In~\cite{botvinik2020variability}, 70 teams analyzed the same task-fMRI dataset, each with their usual pipeline, leading to 70 different analytical conditions. They found substantial differences in the results obtained across teams, in terms of statistic maps but also answers to binary hypotheses. Other studies have explored the variability resulting from specific aspects of the processing and analysis protocol, including operating system~\cite{glatard2015reproducibility}, differences of software and software versions ~\cite{pauli2016exploring,bowring2019exploring,gronenschild2012effects}(which have different implementations to perform various operations), as well as the choices regarding the processing steps applied on the data~\cite{li2021moving}.
The question of whether analytical variability makes it impossible to combine subject data for between-group analyses in these conditions is still widely unanswered. To the best of our knowledge, all of the previous studies focused on how analytical variability affects the reproducibility of existing results in neuroimaging, by using different pipelines to complete a similar analysis in which the processing applied on all subject data is the same, and comparing the results obtained across pipelines. Here instead, we look at the compatibility between data which have been processed differently at subject-level.

As mentioned, we may expect to see a growing proportion of processed data among shared subject data in the future. If we want to use such data, coming from different sources, for between-group studies, it would be necessary to ensure that the differences in processing pipelines used on each subject do not impact the results of between-group analyses and give us something different from what we would have had if all the subject data had been analyzed in the same way. In this study, we focus on how the analytical variability in subject-level pipelines can impact the results of between-group studies.

In order to assess the validity of between-group studies combining data that were processed differently at the subject-level, we carried out a series of between-groups analyses, using data from HCP multi-pipeline dataset\cite{hcp_multi_pipeline}. In each group, we used subject-level data that were processed with different processing pipelines, that implement the same steps and only differ on a set of predefined parameters that allowed us to study individually the impact of each parameter. In the following sections, the term ``pipeline" will be used to designate subject-level pipelines only. We computed the p value $p$ associated to $H_0$: ``no mean difference of activation between groups". Since we used different but similar subjects in each group, by construction, $H_0$ is true, and we expect $p$ to be uniformly distributed, and in particular $P(p<\alpha)=\alpha$.

\section{Material and Methods} \label{matmet}

\begin{figure}[!ht]
\centering
\includegraphics[width=\linewidth]{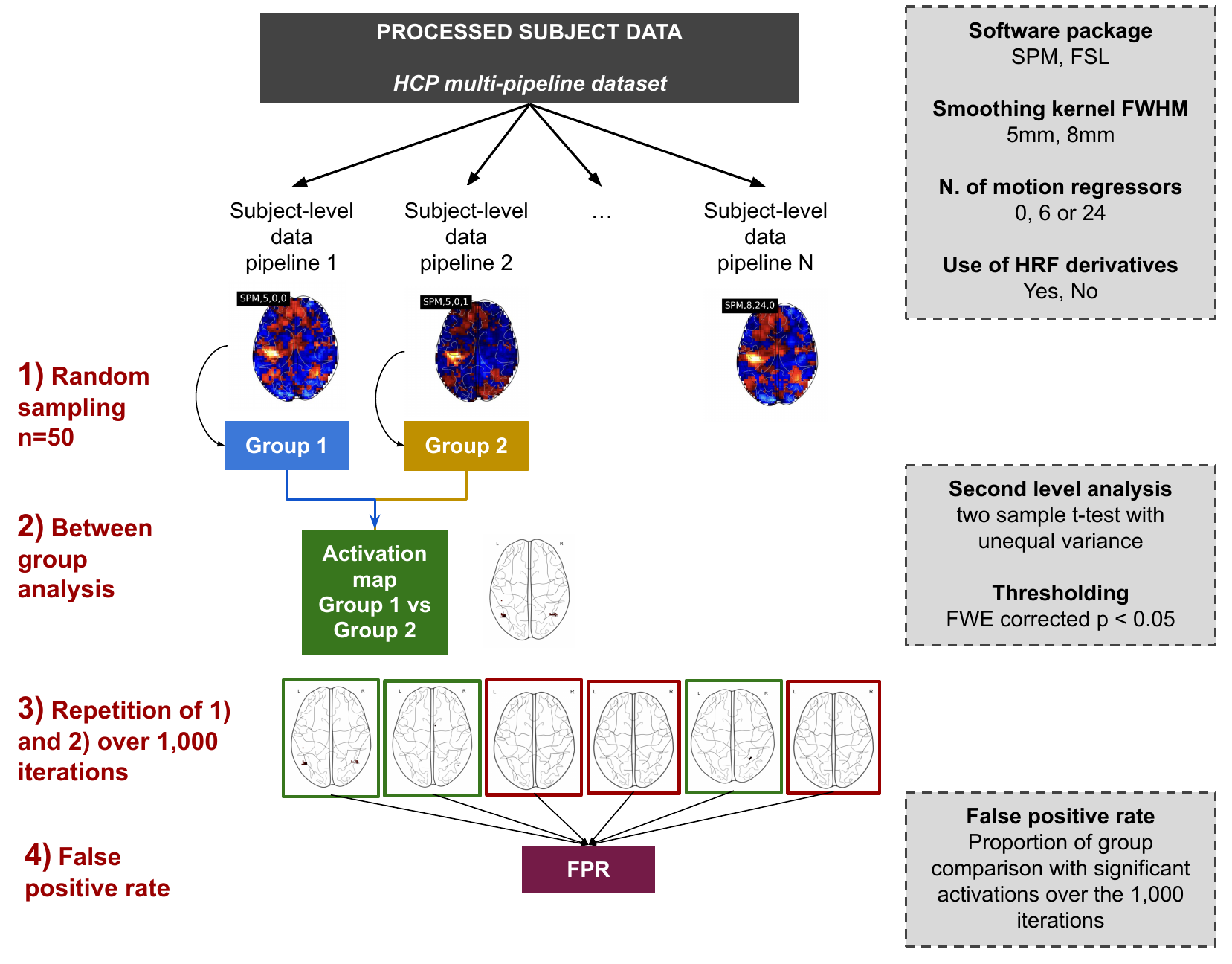}
\caption{Overview of the method: 1) sampling of n=50 subject-level data for each group (i.e. one group = one pipeline) from the HCP multi-pipeline dataset~\cite{hcp_multi_pipeline}, 2) between-group analysis ``Group 1 $>$ Group 2", 3) running 1,000 iterations of 1) and 2), and 4) estimation of the false positive rate.}
\label{stepsfpr}
\end{figure}

This study was performed using derived data from the Human Connectome Project. Written informed consent was obtained from participants and the original study was approved by the Washington University Institutional Review Board. We agreed to the Open Access Data Use Terms available at: https://www.humanconnectome.org/study/hcp-young-adult/document/wu-minn-hcp-consortium-open-access-data-use-terms.

The goal of the study was to test the validity of between-group analyses using subject-level data processed with different pipelines. To this aim, we performed between-group analyses and computed the p value $p$ associated with the null hypothesis $H_0$: ``no between-group difference". Since we used different but similar subjects in each group, by construction, $H_0$ is true, and we expect $p$ to be uniformly distributed, and in particular  $P(p<\alpha)=\alpha$. All significant differences detected were therefore false positives, and we used the detection rate as an estimate of the false positive rate (which under the null hypothesis is expected to be equal to the nominal false positive rate). In this study, we estimated empirical false positive rates obtained when comparing two groups of subjects, each processed with a different pipeline, and observed whether or not they diverged from the nominal false positive rate. Subjects in each group were randomly drawn from the HCP multi-pipeline dataset~\cite{hcp_multi_pipeline}.

The steps performed in order to estimate this false positive rate, which will be detailed in the following subsections, are presented on Figure~\ref{stepsfpr}: first, we selected the raw fMRI data processed through the different pipelines implemented in the HCP multi-pipeline dataset~\cite{hcp_multi_pipeline} (section \ref{hcp-multi}). Then, for each pair of pipelines to compare, we performed a between-group analysis on the first-level analysis results. The two groups corresponded to subject data processed with a different pipeline for each group (section \ref{bw-group}). This group comparison was repeated 1,000 times in order to estimate the empirical false positive rate (section \ref{fpr}).

All the scripts used to perform the study (group-level analysis and false positive rate estimation) are available at [TODO].

\subsection{HCP multi-pipeline dataset} \label{hcp-multi}

Subject-level data were obtained from the HCP multi-pipeline dataset~\cite{hcp_multi_pipeline}. This dataset contains subject-level data from 1,080 subjects of the HCP Young Adult S1200 release~\cite{van2013wu} whose unprocessed functional and structural data for the motor task were analyzed using 24 different pipelines. Details about the subject-level pipelines can be found in~\cite{hcp_multi_pipeline}.

The pipelines implemented in the dataset varied on the following set of parameters:
\begin{itemize}
\item Software package: SPM (Statistical Parametric Mapping, RRID: SCR\_007037)~\cite{penny2011statistical} or FSL (FMRIB Software Library, RRID: SCR\_002823)~\cite{jenkinson2012fsl}.
\item Smoothing kernel: Full-Width at Half-Maximum (FWHM) was equal to either 5mm or 8mm. 
\item Number of motion regressors included in the General Linear Model (GLM) for the first-level analysis: 0, 6 (3~rotations, 3~translations) or 24 (the 6~previous regressors + 6~derivatives and the 12~corresponding squares of regressors).
\item Presence (1) or absence (0) of the derivatives of the Hemodynamic Response Function (HRF) in the GLM for the first-level analysis. Only the temporal derivatives were added in FSL pipelines and both the temporal and dispersion derivatives were for SPM pipelines.
\end{itemize}

In this dataset, pipelines implemented in the different software packages were aligned for some parameters by changing the software package default values. In total, those combinations provided a set of 24~different subject-level pipelines ( 2~software packages $\times$ 2 smoothing kernels $\times$ 3~numbers of motion regressors $\times$ 2~HRF ).

\subsection{Post-processing transforms} \label{post-proc}

Subject-level data obtained with different software packages did not have the same dimension, as standard MNI templates used for normalization are different between FSL and SPM. To be able to perform the between-group analysis, we had to resample subject-level data to a common dimension. We used Nilearn~\cite{abraham_machine_2014} (RRID: SCR\_001362) to resample all subject-level data from all pipelines to the MNI152Asym2009 brain template with a 2mm resolution using continuous interpolation. To remove artefacts caused by this type of interpolation, we computed a brain mask with the intersection of all subject-level brain masks from all pipelines. This mask was also resampled to the MNI brain template using nearest-neighbors interpolation and applied to the resampled subject-level data. In the end, subject-level data from different pipelines had the same dimensions and the same brain mask. 

The unit scale of subject-level data used as input to second-level analyses is also different depending on the software package used (SPM or FSL)~\cite{noauthor_spm_2012}. These subject-level data correspond to so called ``contrast maps", and are supposedly expressed in percent BOLD change. However, in SPM, contrast maps unit is closer to 2-3 times an approximation of percent BOLD change due to the mask used to compute the global mean intensity of the brain and thus, the value of each voxel in percent BOLD change, and to the scaling of regressors. FSL contrast maps also have a different unit from percent BOLD chance because mean brain intensity is scaled to 10,000 in order to increase the number of significant digits. To be able to compare data from pipelines implemented in different software packages, we corrected the contrast maps to obtain something closer to approximate percent BOLD change. Contrast maps from SPM were multiplied by $100/250 = 0.4$, as proposed in~\cite{noauthor_spm_2012}. We did not correct the units of the regression coefficients as this effect was less important in block-designs. Voxels values of FSL contrast maps were multiplied by $100/10,000 = 0.01$ to obtain values in approximate percent BOLD change. 

All between-group analyses were performed on resampled, masked and unit corrected subject-level data. We also run the between-group same-pipeline analyses on the original subject-level data from the HCP multi-pipeline dataset to assess the potential effect of these corrections on estimated false positive rates. Results of these analyses are shown in Supplementary Materials.

\subsection{Between-group analyses} \label{bw-group}

We performed between-group analyses comparing two groups of 50 subjects, where two pipelines were selected and respectively applied to all subjects of each group, for multiple pairs of pipelines. The 50 subjects in each group were randomly sampled without replacement, uniformly among the 1,080 subjects.

Using the contrast maps resulting from subject-level analysis for the selected subjects and pipelines, we looked at the second-level contrast corresponding to the between-group difference in means. We performed a one-tailed two samples t-test with unequal variance using a voxelwise $p<0.05$ FWE-corrected threshold to detect whether there was any significant between-group difference.  

First, we checked the validity of the tests when using the same pipeline for both groups (referred to as ``identical pipeline analysis" in the following). Then, we investigated the compatibility between different pipelines (that we will refer to as ``different pipeline analyses" in the following) and explored the impact of changing one parameter or two parameters between the two pipelines (e.g. smoothing kernels FWHM 5 mm versus 8 mm or smoothing kernels + HRF derivatives FWHM 5mm, no HRF derivatives versus 8mm, use of HRF derivatives). 

In order to have consistent second-level software and analysis conditions for all between-group analyses, all second-level analyses were performed with SPM and unequal variance between groups was specified.

\subsection{False Positive Rates Estimation} \label{fpr}

Under the null hypothesis, there is a theoretical 5\% chance to detect any significant difference in activation for a between-group analysis. Thus, for multiple between-group analyses, the proportion of analyses with at least one detected significant difference should converge towards this 5\% false positive rate.

In order to have a sufficient number of analyses to observe the convergence of false positive rates, each between-group analysis with groups processed with different pipelines was repeated 1,000 times with different groups, to estimate the false positive rates for each pair of pipelines. The empirical detection rate was the proportion of between-group analyses, over the 1,000 repetitions, with at least one significant difference detected between two groups, as shown on Figure~\ref{stepsfpr}. 

\subsection{Statistical distributions and P-P plots} \label{plot}

For a given couple of pipeline, we have 1,000 group analyses, which makes a total of more than 150M voxel values. Since the resulting list of voxels obtained is very large and using it for further observations can be very time-consuming, for each between-group analysis using two given pipelines, on which we wanted to make observations, we only took a random sample of 1,000,000 values from the concatenation of statistical values over the 1,000 corresponding group analyses. 

For a given set of statistical values $x_{i}$ that is expected to follow a certain distribution, we can obtain the associated p-values $1-F(x_{i})$ (where F is the distribution function of the p-values). Since the p-values are expected to follow a uniform distribution on [0,1], for a set of N statistical values, the $k^{th}$ ordered p-value is expected to be equal to $k/(N+1)$ (for our samples of 1,000,000 statistical values, $k/1,000,001$). 

P-P plots are usually used to observe how a given set of statistical values diverge from an expected distribution by plotting, for each $k^{th}$ ordered statistical value, the expected associated p-value on the x-axis and the obtained p-value on the y-axis. Here, we used variants of P-P plots by replacing the p-values by the following:
\begin{itemize}
    \item on the x-axis: $log_{10}(pval_{expected})$ instead of the expected p-value
    \item on the y-axis: $log_{10}(pval_{obtained}) - 
log_{10}(pval_{expected})$ instead of obtained p-value.
\end{itemize}

Doing so allowed us to better observe the tails of our sets of statistical values. High statistical values (right tail of our sample) are associated to low p-values, i.e. to high $log_{10}$ of p-values. Statistical values in the tail are higher than expected when $log_{10}(pval_{obtained}) - log_{10}(pval_{expected}) > 0$, i.e. in case of increased false positive findings, and lower when $log_{10}(pval_{obtained}) - log_{10}(pval_{expected}) < 0$.

Distributions of the statistical values were also observed for various between-group analyses, and compared to the the distribution associated to the Student law with 98 degrees of freedom.

\section{Results} \label{results}

\subsection{Identical pipeline analyses}

Figure~\ref{fprid} shows the false positive rates obtained for all identical pipeline analyses, both for SPM and FSL. For all of these, the false positive rates obtained were below the expected value of 0.05, ranging between 0.012 and 0.028 for SPM and between 0.013 and 0.024 for FSL (see Fig.~\ref{fprid}). 

\begin{figure}[!ht]
\centering
\begin{center}
    {\Large \textbf{SPM}}
        
        \vspace{.3cm}
        
    \renewcommand{\arraystretch}{1.3}
    \begin{tabular}{|c|c|c|c|c|}
    \hline
     & \multicolumn{2}{c|}{Smooth 5mm} & \multicolumn{2}{|c|}{Smooth 8mm} \\
     \hline
     & No derivatives & Derivatives & No derivatives & Derivatives \\
     \hline
    0 motion regressors & 0.012 & 0.013 & 0.016 & 0.023 \\
    \hline
    6 motion regressors & 0.015 & 0.006 & 0.024 & 0.013 \\
    \hline
    24 motion regressors & 0.023 & 0.016 & 0.025 & 0.028 \\
    \hline
    \end{tabular}
    
    \vspace{.6cm}
    
        {\Large \textbf{FSL}}
        
        \vspace{.3cm}
        
    \renewcommand{\arraystretch}{1.3}
    \begin{tabular}{|c|c|c|c|c|}
    \hline
     & \multicolumn{2}{c|}{Smooth 5mm} & \multicolumn{2}{|c|}{Smooth 8mm} \\
     \hline
     & No derivatives & Derivatives & No derivatives & Derivatives \\
     \hline
    0 motion regressors & 0.014 & 0.013 & 0.015 & 0.023 \\
    \hline
    6 motion regressors & 0.018 & 0.014 & 0.018 & 0.018 \\
    \hline
    24 motion regressors & 0.015 & 0.013 & 0.016 & 0.024 \\
    \hline
    \end{tabular}
    
\end{center}
\caption{False positive rates (FPR) obtained for between-analyses with same pipeline for both groups with SPM and FSL, for all possible sets of parameter values for number of motion regressors, smoothing kernel FWHM and presence or absence of HRF temporal derivatives. In every cases, FPR is less than 5\%.}
\label{fprid}
\end{figure}

Results obtained for identical pipeline analyses can be used as a reference to be compared with the results obtained when using different pipelines. False positive rates obtained with original contrast maps (non resampled, masked and corrected) were similar to these ones (see~\ref{supp-table1}). 

\begin{figure}[!ht]
\centering
{\Large \textbf{SPM}}
        
        \vspace{.3cm}
\includegraphics[width=1.\textwidth]{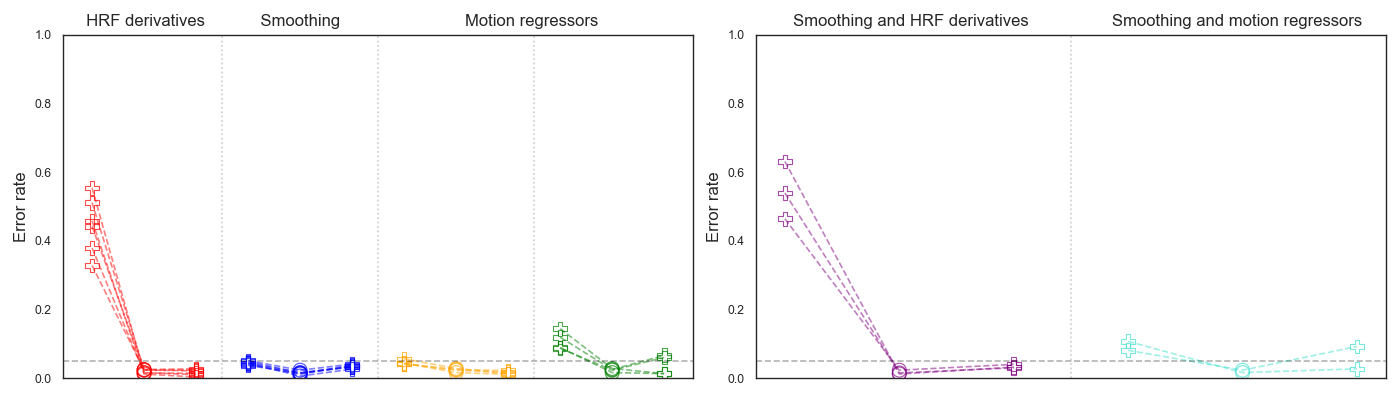}
\vspace{.3cm}
    
        {\Large \textbf{FSL}}
        
        \vspace{.3cm}
\includegraphics[width=1.\textwidth]{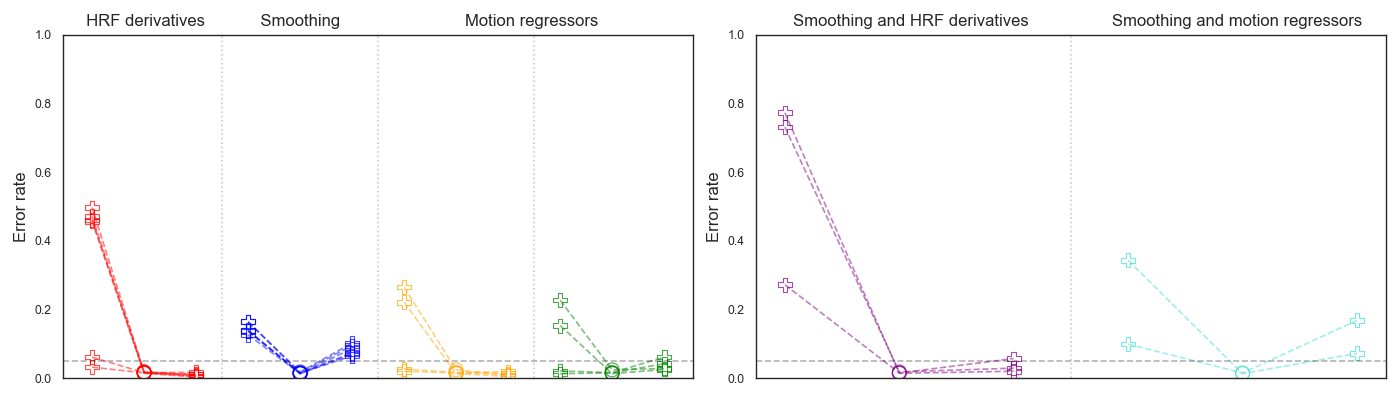}
        
\caption{False positive rates obtained for between-group analyses with different pipelines in SPM (first row) and FSL (second row). Dashed line corresponds to the theoretic $0.05$ threshold. For each software package, we have:
\newline
-First column: for each specific parameter, a group of curves, showing false positive rates obtained for between-group analyses with pipelines which differed only on this parameter, for both left tail (“default $>$ variation”) and right tail (“default $<$ variation”) (cross dots), compared to the false positive rates obtained for identical pipelines analyses with default values for this parameter (round dots)
\newline
-Second column: results for two varying parameters.
\newline
Default values were compared to variations. For one varying parameter: absence of HRF derivatives vs use of HRF derivatives (red), 5 mm vs 8 mm FWHM smoothing kernels (blue) and for the number of regressors, 24 vs 6 (yellow) and 24 vs 0 (green). For two varying parameters, 5 mm FWHM, no HRF derivatives vs 8 mm FWHM, use of derivatives (purple) and 5 mm FWHM, 24 motion regressors vs 8 mm FWHM, 0 motion regressors (turquoise). }
\label{fprcurves}
\end{figure}

\begin{figure}
\vspace{-.5cm}
\centering
{\Large \textbf{SPM}} \\         
\vspace{.1cm}
\includegraphics[width=0.90\textwidth]{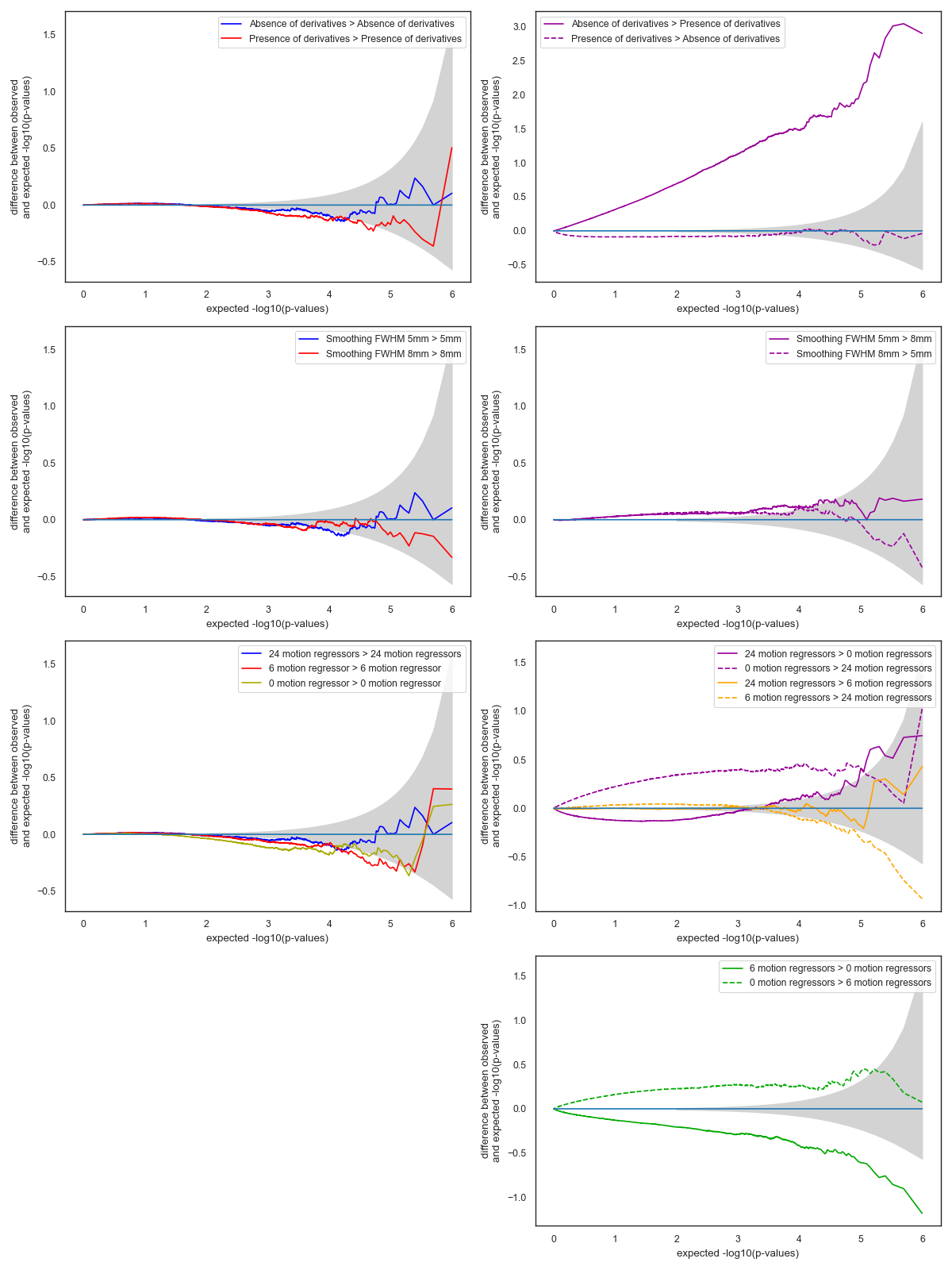}
\caption{Variants of P-P plots for distributions obtained with various analyses under \textbf{SPM}, against the expected distribution, with 0.95 confidence interval, with a \textbf{single} varying parameter for different pipeline analyses. The variations from usual P-P plots are the use of -log(p-values) instead of p-values (to have a more precise observation of the behavior in the right tail) and replacing the obtained -log(p-values) on the y-axis by the difference in obtained and expected -log(p-values).
\newline
Left plots correspond to between-groups analyses with same pipelines in each group. Right plots correspond to between-groups analyses with different pipelines in each group. 
\newline
When not indicated, the default parameters for the pipelines were: 5mm for the smoothing kernel FWHM, 24 motion regressors and no derivatives of the HRF in the GLM. 
\newline
Positive differences indicate invalidity whereas negative differences indicate conservativeness.}
\label{blandaltmanspm1}
\end{figure}

\begin{figure}
\vspace{-.5cm}
\centering
{\Large \textbf{FSL}} \\         
\vspace{.1cm}
\includegraphics[width=0.90\textwidth]{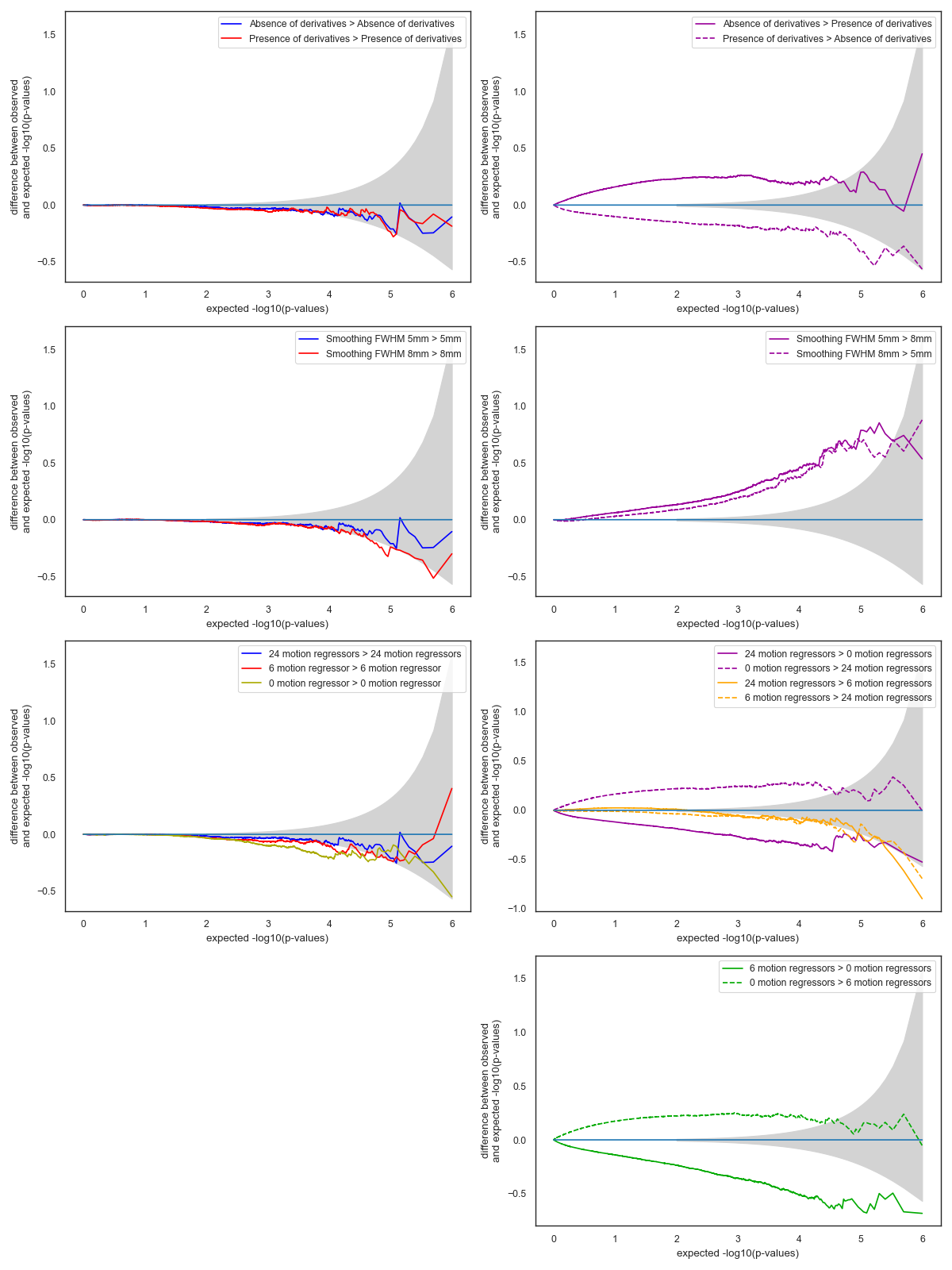}
\caption{Variants of P-P plots for distributions obtained with various analyses under \textbf{FSL}, against the expected distribution, with 0.95 confidence interval, with a \textbf{single} varying parameter for different pipeline analyses. The variations from usual P-P plots are the use of -log(p-values) instead of p-values (to have a more precise observation of the behavior in the right tail) and replacing the obtained -log(p-values) on the y-axis by the difference in obtained and expected -log(p-values).
\newline
Left plots correspond to between-groups analyses with same pipelines in each group. Right plots correspond to between-groups analyses with different pipelines in each group. 
\newline
When not indicated, the default parameters for the pipelines were: 5mm for the smoothing kernel FWHM, 24 motion regressors and no derivatives of the HRF in the GLM.
\newline
Positive differences indicate invalidity whereas negative differences indicate conservativeness.}
\label{blandaltmanfsl1}
\end{figure}

\begin{figure}
\centering
{\Large \textbf{SPM}} \\         
\vspace{.1cm}
\includegraphics[width=0.99\textwidth]{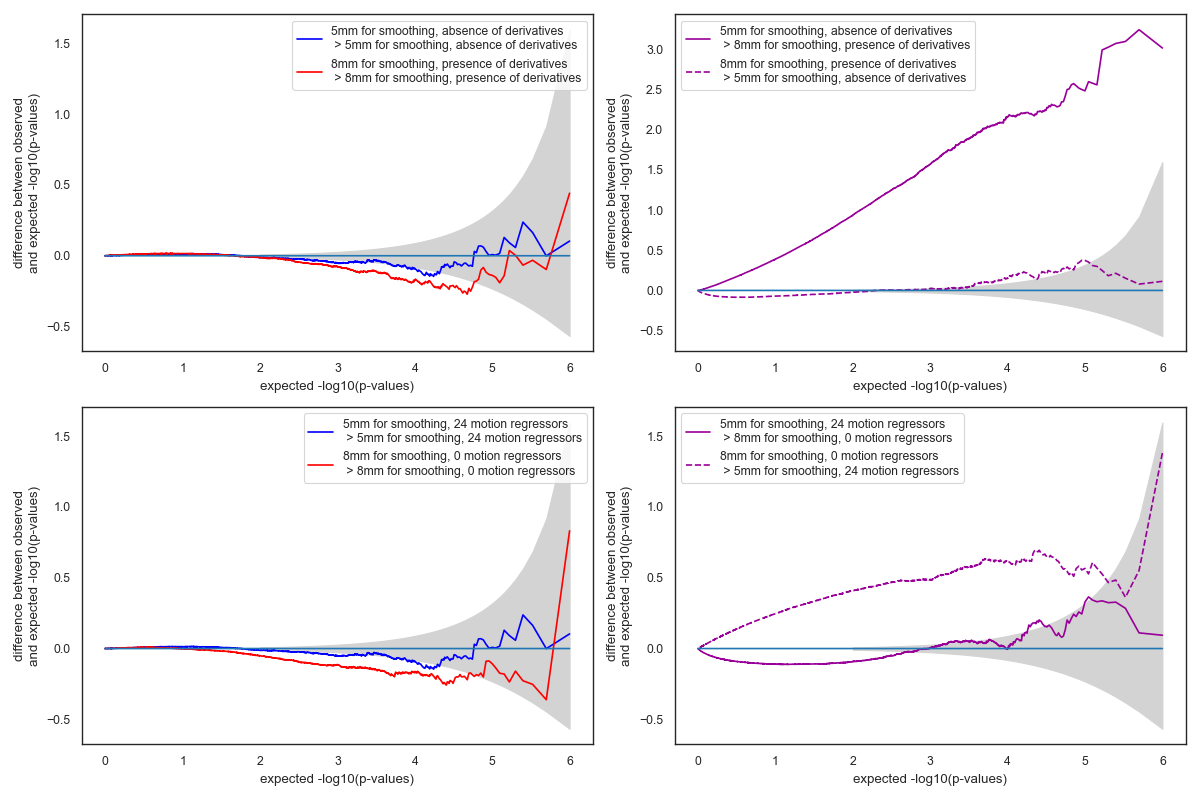}
\caption{Variants of P-P plots for distributions obtained with various analyses under \textbf{SPM}, against the expected distribution, with 0.95 confidence interval, with a \textbf{combination} of varying parameters for different pipelines analyses. The variations from usual P-P plots are the use of -log(p-values) instead of p-values (to have a more precise observation of the behavior in the right tail) and replacing the obtained -log(p-values) on the y-axis by the difference in obtained and expected -log(p-values).
\newline
Left plots correspond to between-groups analyses with same pipelines in each group. Right plots correspond to between-groups analyses with different pipelines in each group. 
\newline
When not indicated, the default parameters for the pipelines were: 5mm for the smoothing kernel FWHM, 24 motion regressors and no temporal derivatives of the HRF in the GLM.
\newline
Positive differences indicate invalidity whereas negative differences indicate conservativeness.
}
\label{blandaltmanspm2}
\end{figure}

\begin{figure}
\centering
{\Large \textbf{FSL}} \\         
\vspace{.1cm}
\includegraphics[width=0.99\textwidth]{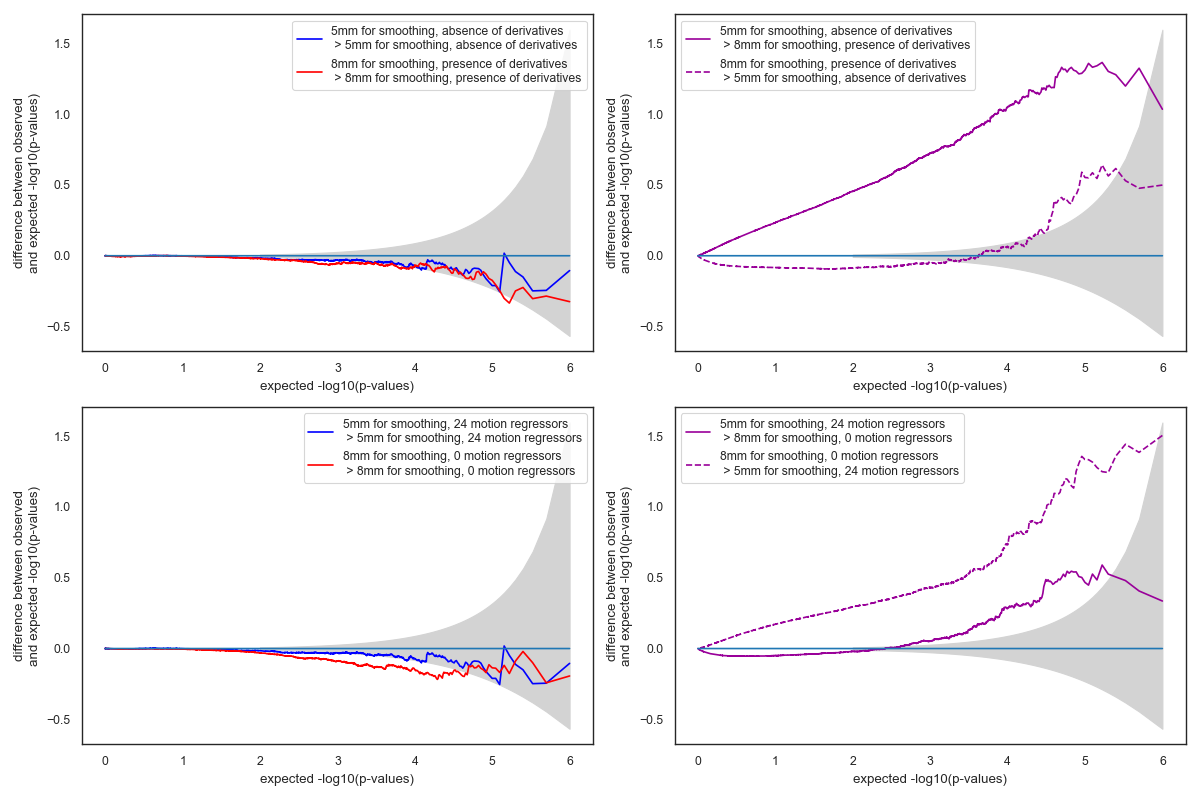}
\caption{Variants of P-P plots for distributions obtained with various analyses under \textbf{FSL}, against the expected distribution, with 0.95 confidence interval, with a \textbf{combination} of varying parameters for different pipelines analyses. The variations from usual P-P plots are the use of -log(p-values) instead of p-values (to have a more precise observation of the behavior in the right tail) and replacing the obtained -log(p-values) on the y-axis by the difference in obtained and expected -log(p-values).
\newline
Left plots correspond to between-groups analyses with same pipelines in each group. Right plots correspond to between-groups analyses with different pipelines in each group. 
\newline
When not indicated, the default parameters for the pipelines were: 5mm for the smoothing kernel FWHM, 24 motion regressors and no temporal derivatives of the HRF in the GLM.
\newline
Positive differences indicate invalidity whereas negative differences indicate conservativeness.
}
\label{blandaltmanfsl2}
\end{figure}

\begin{figure}
    \centering
    {\Large \textbf{SPM}} \\         
\vspace{.1cm}
    \includegraphics[width=\textwidth]{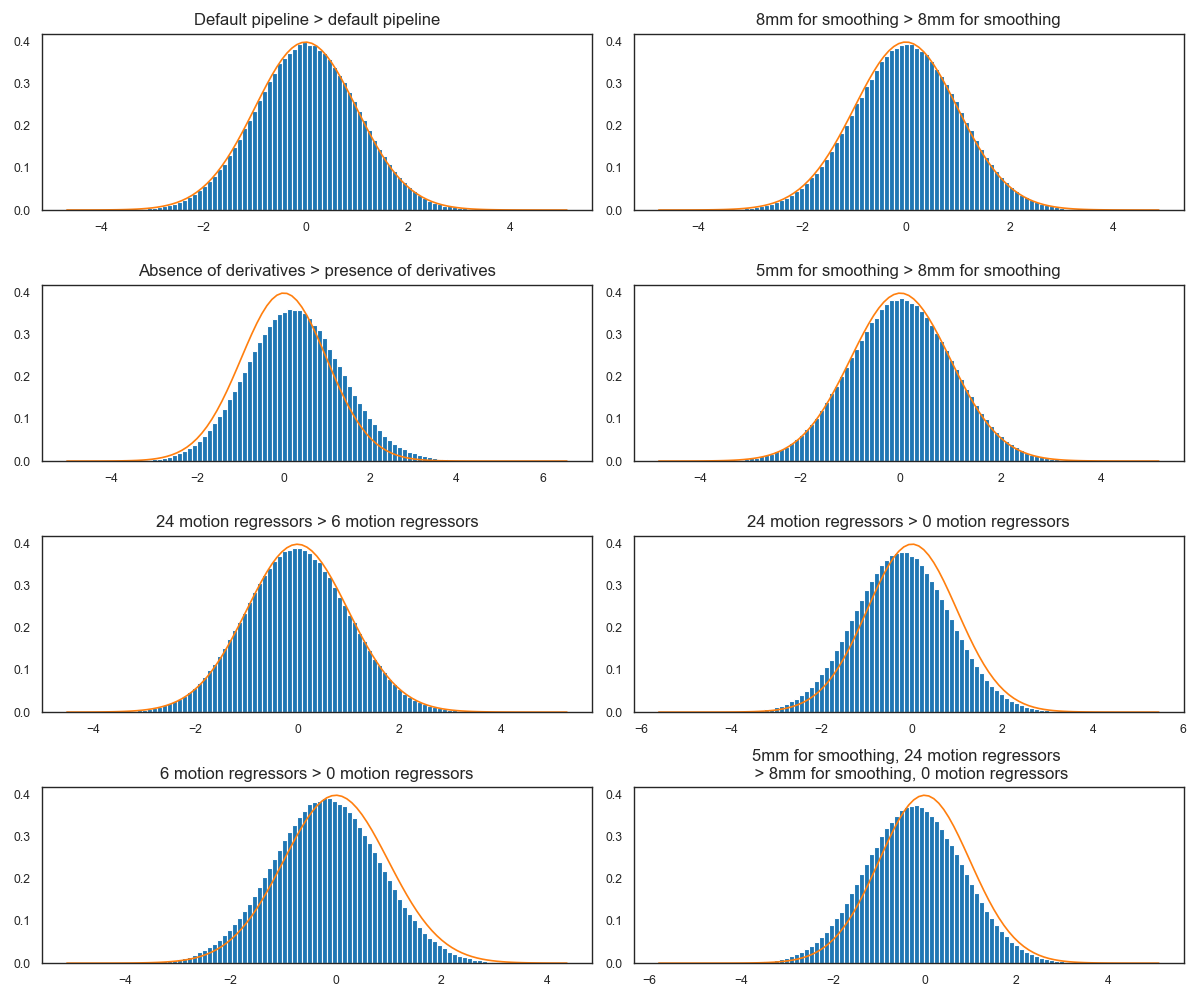}
    \caption{Distribution of statistical values for multiple between-group analyses under SPM, compared to the expected distribution. Pipelines are defined by their differences with the default parameters (5mm for the smoothing kernel FWHM, 24 motion regressors and no temporal derivatives of the HRF in the GLM). Pipelines which differ from the default pipeline are put in bold. The orange curve represents the Student distribution with 98 degrees of freedom, which is the expected distribution in our case under null hypothesis.}
\label{distributionspm}
\end{figure}

\begin{figure}
    \centering
    {\Large \textbf{FSL}} \\         
\vspace{.1cm}
    \includegraphics[width=\textwidth]{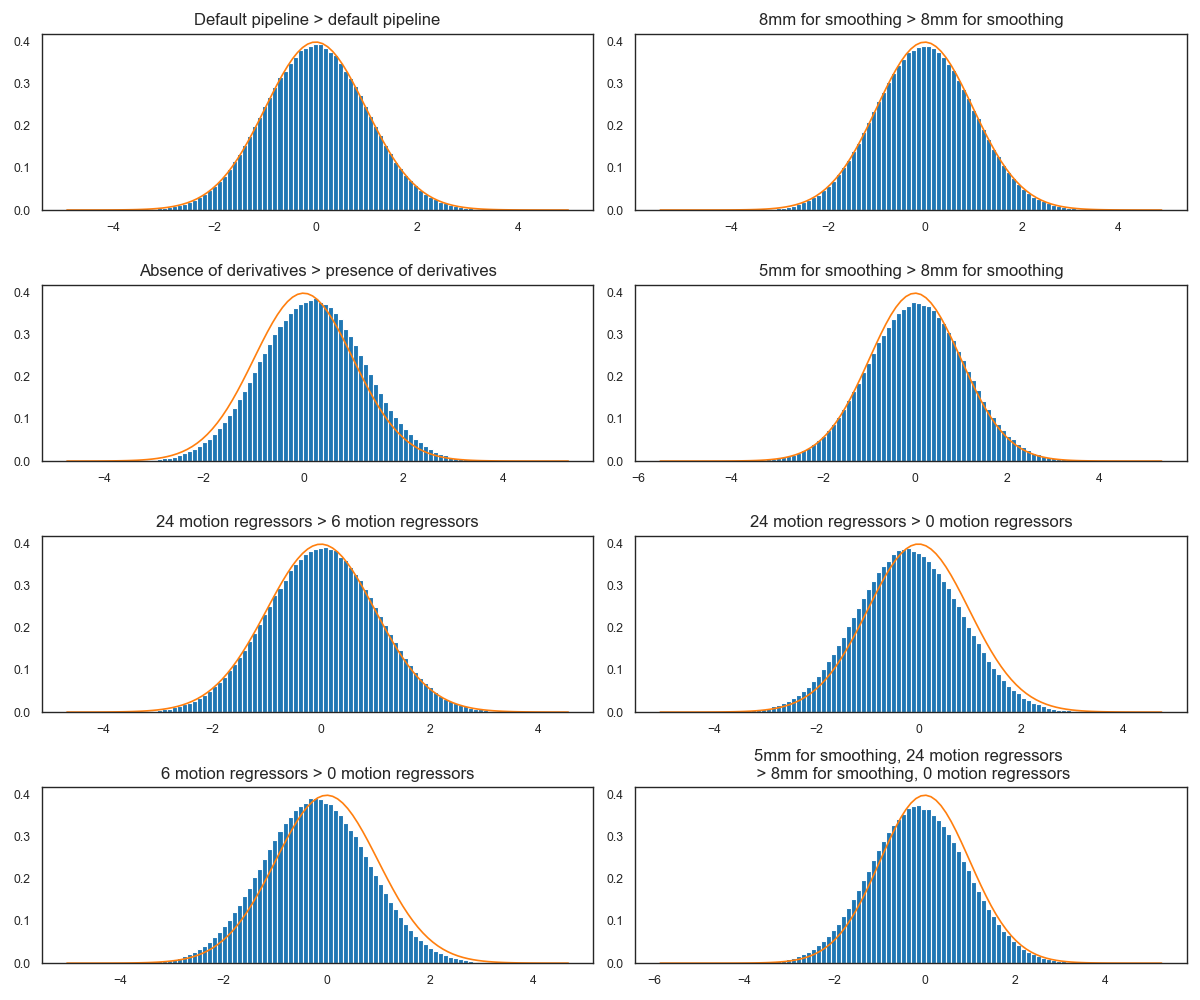}
    \caption{Distribution of statistical values for multiple between-group analyses under FSL, compared to the expected distribution. Pipelines are defined by their differences with the default parameters (5mm for the smoothing kernel FWHM, 24 motion regressors and no temporal derivatives of the HRF in the GLM). Pipelines which differ from the default pipeline are put in bold. The orange curve represents the Student distribution with 98 degrees of freedom, which is the expected distribution in our case under null hypothesis.}
\label{distributionsfsl}
\end{figure}

\subsection{Between-group analyses with different levels of smoothing}

The following subsections explore the results obtained with different pipelines analyses, for which various types of observations were made in each case: false positive rates, statistical distributions and associated P-P plots. For false positive rates, between-group analyses were made on pipelines that differed only on the studied parameter, i.e. all other parameters values were identical between the two pipelines and all combinations were tested (e.g. same software, same n. of motion regressors, same HRF derivatives, different smoothing kernels for studying the effect of smoothing). For P-P plots and statistical distributions, we defined a default value for each parameter: absence of derivatives of the HRF, 5mm for smoothing kernel FWHM value and 24 motion regressors. The results shown on the related figures were obtained with default values for all parameters except for the ones under study: for example, ``5mm $>$ 8mm” refers to ``no HRF derivatives, 5mm, 24 motion regressors $>$ no HRF derivatives, 8mm, 24 motion regressors". We explored the effect of each parameter on the compatibility of subject-level data for each software package separately, as these might be implemented differently in the two software packages. The effect of software package change (``SPM vs FSL") was also explored in the last subsection.

Figure~\ref{fprcurves} (blue curve) shows the false positive rates obtained for the six cases of analyses with different levels of smoothing (in both pipelines, presence or absence of temporal derivatives of the HRF and 0, 6 or 24 motion regressors) with both softwares. For between-group analyses using pipelines with different smoothing --5mm and 8mm-- (blue curves) in both SPM and FSL, the false positive rates were above the 0.05 theoretical false positive rate in FSL and below or close to it in SPM. Compared to identical pipelines analyses, the false positive rates in both directions were all increased and were higher in the direction ``5mm $>$ 8mm" than in the opposite one. This difference between tails is smaller in SPM (false positive rates ranging from 0.037 to 0.051 in the left tail versus from 0.026 to 0.041 in the right tail) than in FSL (ranging from 0.126 to 0.164 versus from 0.07 to 0.102). Observations of the tails of P-P plots (Fig.~\ref{blandaltmanspm1} and~\ref{blandaltmanfsl1}) showed that between-group analyses using pipelines with different smoothing gave invalid results for FSL pipelines in both directions (thus explaining the inflated false positive rates obtained for this combination of parameters). For SPM, values were within the 95\% confidence interval, with only a small positive difference in the direction ``5mm $>$ 8mm", which is consistent with the observed false positive rates.

The behaviors observed on the P-P plots can be explained by the positive shift in mean values and standard deviations observed on the statistical distribution for ``5mm $>$ 8mm" for FSL (Fig.~\ref{distributionsfsl}), which is lower for SPM ((Fig.~\ref{distributionspm})). This effect caused an increase of statistical values, which explains the invalidity of the false positive rates. 

\subsection{Between-group analyses with different number of motion regressors}

To explore the impact of using different number of motion regressors on the compatibility of subject-level results, we performed several between-group analyses with different numbers of regressors. We defined the default value as ``24 motion regressors" and studied the combinations ``24 motion regressors versus 0 motion regressors" and ``24 motion regressors versus 6 motion regressors" (Figure~\ref{fprcurves}, yellow and green curves). The false positive rates obtained for these comparisons were variable across fixed parameters values. Overall, false positive rates were increased compared to identical pipelines analyses, in particular for the left tails (``default > variation"), and were more elevated in FSL than in SPM. The left tails in Figure~\ref{fprcurves} were higher than the right ones, meaning that the comparisons ``24 motion regressors > 0/6 motion regressors" were leading to more false positives than the opposite ones. These observations were valid for all except two comparisons: FSL, 5mm and 8mm smoothing kernel FWHM, no HRF derivatives for both 24 motion regressors versus 0 motion regressors. For these, false positive rates were smaller in the left tails and close to the identical pipelines ones. The P-P plot (Figure~\ref{blandaltmanfsl1}) for the distribution obtained for the combination ``FSL, 5mm smoothing kernel, no HRF derivatives, 24 motion regressors versus 0 motion regressors" confirmed this observation: we found conservative results for the comparison ``24 motion regressors > 0 motion regressors" (full line) and invalid ones for the opposite (dashed line).

For the same comparisons (same values for fixed parameters), false positive rates obtained for ``24 motion regressors $>$ 0 motion regressors" were higher than those obtained for ``24 motion regressors $>$ 6 motion regressors", in both SPM and FSL. This is also noticeable in the P-P plots (Figures~\ref{blandaltmanspm1} and~\ref{blandaltmanfsl1}) in which the comparisons for ``24 motions regressors versus 0 motion regressors" gave more invalid results than the ``24 motions regressors versus 6 motion regressors" ones. Statistical distributions (Figures~\ref{distributionspm} and~\ref{distributionsfsl}) also show a shift in mean and variance for this comparison with 0 motion regressors, for both SPM and FSL. This shift is not as important for the comparison ``24 motion regressors versus 6 motion regressors". The comparison ``6 motion regressors versus 0 motion regressors" was also showed for comparison, and showed similar results as the ``24 motion regressors versus 0 motion regressors" comparison.

\subsection{Between-group analyses with different HRF models}

Adding a temporal derivative to model the HRF was the most impacting of all three varying factors in both software packages. Figure~\ref{fprcurves} shows the false positive rates obtained for different pipeline analyses with different models of the HRF (red curves). In each case, the left tail (``no HRF derivatives $>$ use HRF derivatives") was showing higher false positive rates, above the 0.05 threshold, than the right one (around 0.5 for the left tails compared to 0.02 for the right ones for both SPM and FSL). These values for the left tails were the largest ones obtained for between-group analyses with only one varying parameter. 

Similarly to what we observed for analyses with different motion regressors, two combinations gave valid false positive rates on the left tail in FSL: 5mm or 8mm smoothing kernel FWHM, 24 motion regressors and no HRF derivatives vs use of HRF derivatives (default parameters). These two analyses led to values around the 0.05 threshold (0.032 and 0.061 respectively for the left tail), similarly to what we observed for the comparison of FSL, 5mm or 8mm smoothing kernel FWHM, no HRF derivatives, 24 motion regressors versus 0 motion regressors. We can suppose that in FSL, when using 24 motion regressors and no HRF derivatives, a change in either number of motion regressors or use of HRF derivatives has a lower impact on statistic maps and thus, on the compatibility of subject-level data. 

Figures~\ref{blandaltmanspm1} and~\ref{blandaltmanfsl1} show the P-P plots obtained for the default parameter comparison in SPM and FSL. In SPM, values were far from the 95\% confidence interval (grey area) for the comparison ``absence of derivatives $>$ presence of derivatives". For FSL, values were closer to the valid area, even if these were still outside the 95\% confidence interval. Same observations could be made on the statistical distributions for both SPM and FSL (Figures~\ref{distributionspm} and~\ref{distributionsfsl}): both showed a shift in mean and variance, but this was smaller for FSL, probably due to the specific combination (default parameters) studied. We observed the statistical distribution and P-P plots for a different combination of FSL pipelines with different use of HRF derivatives in Supplementary Materials~\ref{fig:supp-fig1} and found similar results as those observed for  SPM.

\subsection{Combined effects of parameters}

We observed the combined effects of:
\begin{itemize}
    \item differences in smoothing and HRF model
    \item differences in smoothing and motion regressors
\end{itemize}

For the first set of between-group analyses (5mm, no derivatives $>$ 8mm, presence of derivatives), under SPM, the results were close to those obtained for the analyses ``no HRF derivatives $>$ use of HRF derivatives" (from 0.46 to 0.63 in the combined effect analysis and from 0.32 to 0.52 in the exploration of HRF derivatives effect only) (Fig.~\ref{fprcurves}). The effect of the smoothing kernel FWHM was not very important in the isolated analyses ``5mm vs 8mm smoothing kernel FWHM" in SPM, which might explain the low increase of false positive rates in the combined effect analyses. Under FSL, the previous analyses on the effect of each of these parameters separately both gave increased false positive rates, and their combined effect largely increased the false positive rates (up to 0.77) compared to the effect of HRF derivatives alone (up to 0.49). For both SPM and FSL, the comparison ``5mm, no HRF derivatives $>$ 8mm, use of HRF derivatives" (left tails) showed higher false positive rates than the opposite one, which is consistent with the observations made on the effect of HRF derivatives alone. Similar observations can be made on the P-P plots on Figure~\ref{blandaltmanfsl2} and~\ref{blandaltmanspm2}. 

For the second set of analyses (5mm, 24 motion regressors $>$ 8mm, no motion regressors), observations were similar, but false positive rates were lower than when studying a combined effect with changes in HRF derivatives. In SPM, we found similar values in the combined effect analysis and in the effect of motion regressors (24 versus 0 motion regressors) alone, with no effect of adding differences in smoothing kernel FWHM values. In FSL, values were slightly increased in the combined effect compared to motion regressors alone, probably due to the combination with smoothing effect which already led to increased false positive rates. For both SPM and FSL however, we observed shifts in the distributions of statistical values (Figure~\ref{distributionspm} and~\ref{distributionsfsl}). These shifts were similar to those obtained for changes in motion regressors only. 

\subsection{Between-group analyses with different software packages}

\begin{figure}
    \centering
    \includegraphics[width=8cm]{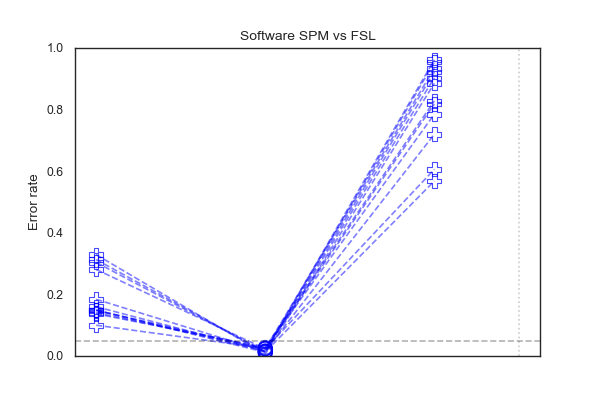}   \caption{False positive rates obtained for between-group analyses with different software packages. Dashed line corresponds to the theoretic $0.05$ threshold.
    \newline
    For each specific parameter, a group of curves, showing false positive rates obtained for between-group analyses with pipelines which differed only on this parameter, for both left tail (“default $>$ variation”) and right tail (“default $<$ variation”) (cross dots), compared to the false positive rates obtained for identical pipelines analyses with default values for this parameter (round dots).}
    \label{fpr-intersoft}
\end{figure}

\begin{figure}
    \centering
    \includegraphics[width=16cm]{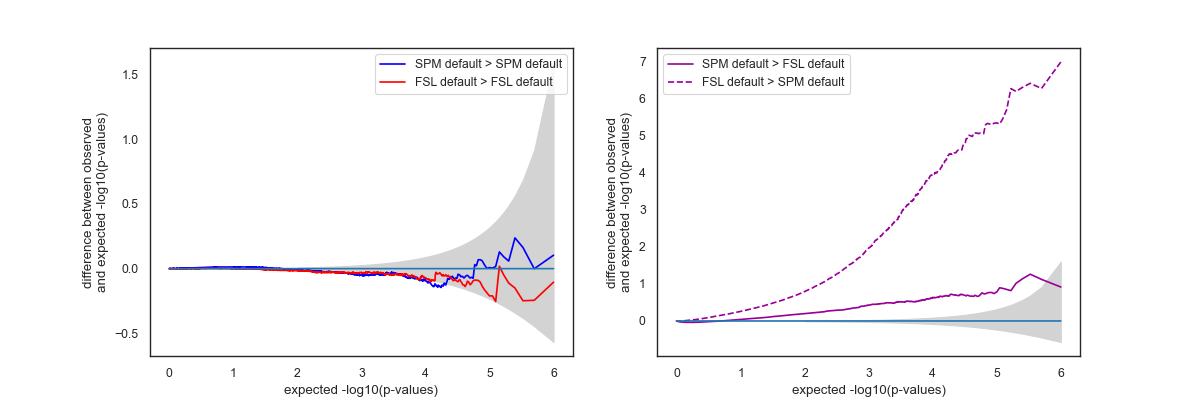}
    \caption{Variants of P-P plots for distributions obtained with between-software analyses. 
    \newline
    Left plots correspond to between-groups analyses with same pipelines in each group. Right plots correspond to between-groups analyses with different pipelines in each group. 
    \newline
    When not indicated, the default parameters for the pipelines were: 5mm for the smoothing kernel FWHM, 24 motion regressors and no temporal derivatives of the HRF in the GLM.
    \newline
    Positive differences indicate invalidity whereas negative differences indicate conservativeness.}
    \label{ppplot-intersoft}
\end{figure}

\begin{figure}
    \centering
    \includegraphics[width=8cm]{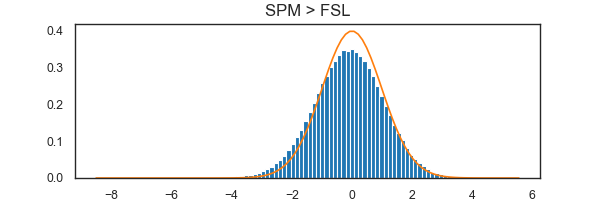}
    \caption{Distribution of statistical values for between-software analyses, compared to the expected distribution.}
    \label{distrib-intersoft}
\end{figure}

We also explored the compatibility of subject-level data obtained with different software packages, here FSL and SPM. We performed the analyses for all possible combinations SPM vs FSL: 2 smoothing kernels $\times$ 3 numbers of motion regressors $\times$ 2 HRF, corresponding to 12 between-software comparisons. All false positive rates are displayed in Figure~\ref{fpr-intersoft}. For these between-software analyses, false positive rates were largely increased compared to identical pipelines analyses and highly above the 0.05 theoretical false positive rates. For the direction ``SPM $<$ FSL", we obtained higher values than in the opposite direction (from 0.1 to 0.32 in the direction ``SPM $<$ FSL" and from 0.56 to 0.95 for ``SPM $>$ FSL"). This observation was consistent with the P-P plot, which showed a large deviation from the 95\% confidence interval for the direction ``SPM > FSL" (Figure~\ref{ppplot-intersoft}). Figure~\ref{distrib-intersoft} shows the distribution of statistical values for the between-software comparison with all other parameters set with default values (i.e. 5mm smoothing kernel, 24 motion regressors and no HRF derivatives). We can see a large shift in terms of mean and standard deviation of values. This shift was larger than those observed, for instance, for the effect of HRF derivatives, which was the most impacting factor on within-software comparisons. 

\section{Discussions}

We observed that, when combining data from different pipelines, the effect of differences in subject-level processing depended on which parameter differed between the pipelines. Some differences in parameters gave conservative results, same as what we observed for identical pipelines analyses, while other differences gave invalid results. For this reason, there are no guarantees regarding the validity of positive results that could be obtained, in practice, when performing between-group analyses under similar circumstances, using subject data processed differently. If significant differences are observed in experiments under these conditions, it cannot be determined whether these differences are associated to an existing effect or to the differences in processing.

Our results also showed that when performing analyses using the same pipeline on all subject data (as traditionally done in the existing literature), results were valid for all analyses. For this reason, the invalid results, obtained when combining subject data processed differently, suggest that it is necessary to consider how analytical variability may affect the results when combining data, and not that a specific subject-level pipeline is responsible for the invalidity of the results and should not be used in any case. Although the false positive rates obtained in this situation were lower than the expected 5\% rate, the results were similar to those obtained for a similar framework in~\cite{eklund2016cluster}.

Our results for different pipeline analyses suggest that some factors have a larger impact on data compatibility than others. We saw that for differences regarding the size of the smoothing kernel and number of motion regressors (6 versus 24 motion regressors) with SPM software package, results were similar to those obtained with identical pipeline analyses, suggesting that subject data can be combined without having to consider the differences in pipelines, if this is the only difference. This is not the case for differences in the use of HRF derivatives and use of motion regressors (no motion regressors versus 6 or 24 motion regressors), which gave invalid results. We also saw that combining multiple differences in parameters could result in bigger effects, depending on the effect of each parameter alone. The combination of two parameters that both have a high effect on compatibility led in our case to inflated false positive rates, while the combination of low-effect parameters did not led to higher false positive rates (e.g. smoothing and motion regressors in SPM). Our observations of the distributions of statistical values also allowed us to have an insight at how each difference in parameter affected the results (differences in mean and variance compared to the expected distribution), as well as their combination.

Some comparisons showed different results depending on the values of the non-varying parameters: for example, all analyses using pipelines with the same software package, the same number of motion regressors, the same use of HRF derivatives and different levels of smoothing (5mm and 8mm) gave a similar inflation of false positive rates, for all possible cases of numbers of motion regressors and models of the HRF (identical in both pipelines). This, and the effects observed with combined variations of parameters, suggests that it may be possible to model the effect caused by specific variations in the subject-level pipelines. To do this modelisation with processed subject data taken from existing datasets, the pipelines used on the data have to be shared in a format that would allow to know and reproduce the exact processing applied on the data.

However, the ability to model this effect is limited to specific variations. Indeed, for each variation of parameter between pipelines, we saw inconsistent effects across the two software packages under study (SPM and FSL). Overall, observations were similar, but false positive rates were increased in FSL compared to SPM for the same comparison. Specific combinations of parameters in FSL also led to lower false positive rates than others for the same comparison. This suggests that some parameters values are more robust to changes when combined together, here, in FSL, when using 24 motion regressors and no HRF derivatives, combining data in which HRF derivatives were used or no motion regressors led to false positive rates closed to the identical pipelines ones. 

The most important source of invalidity was found when studying the effect of differences in software packages. SPM and FSL both implement similar pipeline steps with different settings. While we tried to align some parameters between the two software packages by changing the software package default values (e.g. smoothing kernel, type of HRF, etc.), some steps are specific to each software and cannot be changed by the user, causing potential differences between the results. We tried to correct some of these differences, in particular for the unit scale of subject-level contrast maps. But, even with these corrections, we still found highly elevated false positive rates when comparing pipelines with the same values for the parameters under study and different software packages. We suppose that software packages differences are too important and too little understood to consider combining subject-level data from two different software packages. Performing these combinations might request deeper analyses to correct for all potential differences.

While our study shows that invalid results can be obtained when performing between-group analyses where each subject data is processed with a different subject-level pipeline depending on the group, this is not the only type of situation which may appear where there are combinations of subjects data processed differently. Subject data from multiple subject-level pipelines may also be combined within a group, and multiple groups may contain data processed with a same subject-level pipeline.

The situation that we have here, where subject-level pipelines differ depending on the group, may occur for example when using data from various datasets, if the subject data within each group comes from the same dataset. For example, specific datasets have been created to study various neurological disorders (Alzheimer's Disease Neuroimaging Initiative (ADNI)~\cite{jack2008alzheimer} for Alzheimer's disease, Autism Brain Imaging Data Exchange (ABIDE)~\cite{di2014autism} for autism, etc), and researchers may want to use them to compare groups of subjects where each group corresponds to a specific disease. If the datasets they want to use for these comparisons only include processed subject data, it is likely that, within each dataset, processing applied on the subject data will be the same, and that across datasets, processing will differ. Each dataset may then be associated to a specific processing pipeline for its subject data and consequently, each group also. In this situation, in practice, using the same pipelines that we used here will have an effect which is similar to the one observed here. Also, since the methodological choices made to create the pipelines that we used (in terms of steps performed and chosen parameter values) are very common in the literature~\cite{carp2012plurality}, the resulting pipelines are typical examples of pipelines used in task-fMRI.

We chose to study variations induced by 4 types of parameters (software package, HRF, smoothing, motion regressors), within each software package, but there are many more used in practical conditions: researchers might use different software versions, perform or not specific substeps within the analysis (for example, slice-timing correction), using different models of the HRF, etc. In real conditions, if researchers combine subject data processed differently, the differences between pipelines will likely be more important.

In future works, other analyses may be done for other varying parameters such as the ones mentioned above, with the same framework that we used in this study. Doing so would provide a more precise knowledge of the real extent of the impact of analytical variability, and maybe lead to observations of phenomenons that are different from what we observed here. For example, even though our study shows no evidence that this is the case, differences in subject-level pipelines with an effect which would be the opposite of what we have seen here (lower detection than with identical pipelines) may exist: in this case, the differences in subject-level pipelines could also make it harder to detect existing effects. 

We have seen that the effects of analytical variability often prevent us from combining data without considering the differences in processing pipelines. Our observations on the data suggest that this effect may be corrected, and thus it may be possible to overcome the problem of analytical variability when combining data. For other sources of variability, methods have been proposed to correct this effect: for example, correcting the variability resulting from imaging site and scanner effect (technical variability) in neuroimaging~\cite{beer2020longitudinal,fortin2016removing}. Finding solutions for issues related to analytical variability in general is a growing research subject in many fields: for example, finding a consensus between different results obtained with different methods, which would correspond to results obtained with different complete processing pipelines~\cite{dafflon2020neuroimaging}). Our study has already allowed us to measure the impact of analytical variability on the combination of data processed differently for a set of specific varying parameters, and gives a framework for future studies to measure it for other parameters. The ability to combine processed data easily would be a major step toward more reproducible research, and the correction of the effect of analytical variability in such a situation is still a widely unanswered research question. We could for instance benefit from the knowledge of the machine learning community, in which such a problem could be seen as a style transfer issue, and could be solved with techniques such as image-to-image translation~\cite{pang_image--image_2021}. Therefore, investigating ways to solve this issue may become a topic of increasing interest in the near future for the neuroimaging community. 

\subsection{Conclusion}

Our study shows that, when combining subject data which have been processed differently, the validity of the results obtained depends on the differences between the pipelines used on subject data. While there are parameters for which differences in values between pipelines do not seem to have major effects, some other parameter differences can produce invalid results, suggesting that it is impossible to combine processed fMRI data without taking into account these differences in subject-level processing. In further works, we will develop methods to correct the effect of analytical variability that we observed in this situation.

\subsubsection{Acknowledgements} Data were provided by the Human Connectome Project, WU-Minn Consortium (Principal Investigators: David Van Essen and Kamil Ugurbil; 1U54MH091657) funded by the 16 NIH Institutes and Centers that support the NIH Blueprint for Neuroscience Research; and by the McDonnell Center for Systems Neuroscience at Washington University.

Xavier Rolland was supported by Region Bretagne (ARED Varanasi) and by EU H2020 project OpenAIRE-Connect (Grant agreement ID: 731011). Elodie Germani was supported by Region Bretagne (ARED MAPIS) and Agence Nationale pour la Recherche for the programm of doctoral contracts in artificial intelligence (project ANR-20-THIA-0018). 

\bibliography{references}

\end{document}